\input{aipcheck}

\documentclass[
    ,final            
  ]
  {aipproc}
\layoutstyle{6x9}

\begin{document}

\title{Pulsar electrodynamics: a time-dependent view}


\author{Anatoly Spitkovsky}{
  address={Kavli Institute for Particle Astrophysics and Cosmology, \\
Stanford University, PO Box 20450, MS 29, Stanford, CA 94309}
}

\begin{abstract}
Pulsar spindown forms a reliable yet enigmatic prototype for the energy loss processes in many astrophysical objects including accretion disks and back holes. In this paper we review the physics of pulsar magnetospheres, concentrating on recent developmen
ts in force-free modeling of the magnetospheric structure. In particular, we discuss a new method for solving the equations of time-dependent force-free relativistic MHD in application to pulsars. This method allows to dynamically study the formation of t
he magnetosphere and its response to perturbations, opening a qualitatively new window on pulsar phenomena. Applications of the method to other magnetized rotators, such as magnetars and accretion disks, are also discussed. 
\end{abstract}

\maketitle


\section{Introduction}
Pulsars are the prototypes for a class of astrophysical objects that emit most of their spindown energy in the form of electromagnetic flux. Such objects typically consist of a central conducting body surrounded by a highly magnetized rotating magnetosphe
re and a relativistic outflow. Understanding the structure and inner workings of pulsar magnetospheres is therefore important not only for interpreting the vast and constantly growing collection of observations of pulsars and their winds, but also as a keystone to understanding the pro
cesses in the magnetospheres of other magnetically dominated systems, such as magnetars and accretion disks. 
While the basic features of pulsar magnetospheres were understood for a long while, the road to constructing a self-consistent picture had been difficult. The main obstacle is the nonlinear nature of plasma dynamics in the presence of strong electromagnet
ic fields. This inhibited the progress in obtaining analytical solutions, and only in recent years the subject gained new interest due to the development of numerical techniques able to deal with the nonlinearities and singularities. In this paper we will
 first review the basics of unipolar induction, discuss the issue of plasma production in the magnetosphere, and then concentrate on the magnetospheric solutions in the limit of force-free magnetohydrodynamics, underlining both the physical properties of 
the solutions and the numerical issues involved. 

\section{Pulsar spindown}
The underlying mechanism for converting rotational energy into electromagnetic energy loss from a pulsar is the unipolar induction from a rotating magnetized sphere. Rotation at an angular frequency $\Omega$ of a conducting sphere in a magnetic field gene
rates potential difference between the pole and the equator $V\sim \Omega \Psi$, where $\Psi$ is the enclosed magnetic flux. This potential is due to the Lorentz force separating the charges inside of a rotating conductor. For Crab pulsar with $B \sim 10^
{12}$G this potential is $10^{16}$V , and for a typical AGN accretion disk with $B \sim 10^{4}$ G the potential is $10^{20}$V . If this potential is harnessed to drive a current $I$, the system will lose energy at the rate $W=IV=V^2/Z$, where $Z$ is the i
mpedance of the load, usually being the impedance of free space in astrophysics. For the Crab pulsar we then estimate the energy loss of $10^{38}$ergs/s, and for an AGN disk -- $10^{46}$ergs/s. Such numbers are close to the characteristic observed values,
 reinforcing the idea that rotating magnetized objects loose energy electromagnetically. How the currents are driven in a pulsar and what is the actual circuit remains to be understood in detail, but the basic idea is as follows.  
Potential distribution due to unipolar induction leads to external electric fields around the sphere (for a review see \cite{MichelLi99}), with a large component along the magnetic field. These accelerating fields cause emission of charges from the surfac
e \cite{GJ}. Later evolution depends on whether there is a source of extra plasma in the magnetosphere (pair production), or all plasma has to come from the surface. 

When the plasma is emitted only from the surface, the magnetosphere is charge separated (only one sign of charge is present at a given location), and is subject to interesting dynamics of charge separated flows. Initially, particles are accelerated off th
e surface in the space charge limited flow and a small fraction of charges can get accelerated to near the full potential of the star, decoupling from the magnetic field lines near the light cylinder and leaving to infinity. However, most of the charges e
xtracted by the quadrupolar electric field remain on magnetic field lines. The positive charges extracted in the equatorial region of the star have no way to leave the system and accumulate in a disk near the equator. Negative charges form clouds (or "dom
es") near the poles \cite{KPM84}. The electrostatics of this external charge distribution is such that the accelerating field near the star is reduced as the cloud accumulates, and within one rotation of the star the system comes to an equilibrium with do
mes and disks of charge hanging around the star and preventing additional injection of plasma. There is some residual injection to compensate for the charges that return to the star, but overall the system reaches equilibrium. The magnetosphere is not com
pletely filled with charge, as the disk-dome configuration extends to only several radii of the star. Further evolution depends on the dimensionality of the problem. In 2D axisymmetric setup the disk-dome configuration is stationary and stable to perturba
tions, including perturbations due to the intermittent injection of neutral plasma \cite{SMT01}. However, in 3D the disk-dome system is unstable to the diocotron instability \cite{SA02, S04}. Because plasma does not fill the full magnetosphere, the charge
 density deviates from the Goldreich-Julian density $\rho_{GJ}=\Omega \cdot B/4 \pi e c$ \cite{GJ} that is required for corotation of the magnetosphere with the star. Therefore, the charge-separated "electrosphere" rotates differentially. This is possible
 because unlike in MHD the field lines are not equipotentials and can cross regions of vacuum before passing through a charge distribution. This way even though the star is rotating, not all field lines are "rotating" with the same angular velocity (motio
n of the field is only well defined when it is immersed in plasma, then the velocity of the fieldline is the $E\times B$ drift velocity). The differential rotation of the plasma around a neutron star is providing the free energy for the growth of the dioc
otron instability, which is an analog of the Kelvin-Helmholtz instability in hydrodynamics. The results of the time-dependent simulations using particle-in-cell (PIC) technique are presented in \cite{SA02, S04} and the analytic instability analysis can be
 found in \cite{Petri02b}. The diocotron instability leads to radial transfer of charge in the equatorial disk perpendicular to the magnetic field lines. This happens due to the azimuthal component of electric field that develops during this nonaxisymmetr
ic instability. In the simulations the redistribution of charge leads to the approach to corotation in the inner magnetosphere, while the charge cloud is growing to larger radii. Eventually the cloud reaches the light cylinder, and the nature of the insta
bility switches from  electrostatic to electromagnetic -- the currents induced by the rotation of the charges start to modify the poloidal magnetic field. While in principle the PIC method takes this into account, the present simulations have not been run
 long enough with sufficient precision to resolve this transition. Before more simulation work is invested into the problem of filling the magnetosphere with charge, it is not clear whether the charge-separated magnetosphere, as in the original work by Go
ldriech and Julian \cite{GJ}, is viable for describing the behavior of pulsar magnetosphere. While the simulations are pointing towards the relaxation of space charge configuration to corotation with the star, it is not clear how this solution jump-starts
 the flow of currents through the magnetosphere. Effects of the oblique magnetosphere introduce additional wave pressure near the light cylinder from the oscillating magnetic dipole \cite{S04}. This pressure drives some charges away, but it remains to be 
understood whether obliqueness is crucial for initiating the current flow. Besides pure academic interest, models of such "dead" pulsars are relevant for the neutron stars at late stages of evolution when pair formation in the magnetosphere is suppressed,
 or to the isolated neutron stars that do not readily form pairs due to the very large magnetic field at the surface. 
 
\section{Force-free electrodynamics}

While charge-separated models of the pulsar magnetosphere were the first in the literature \cite{GJ}, more attention was paid over the years to the opposite limit, when the plasma filling the magnetosphere is quasineutral. Physically, the quasineutral pla
sma is likely to come from pair production in the vacuum gaps near the surface due to curvature radiation photons from the primary particles or from inverse-Compton scattered thermal photons from the surface. When sufficient plasma is available to provide
 flux-freezing, relativistic MHD can be used to describe the magnetosphere. In fact, due to magnetic energy domination over plasma energy, \emph{force-free} MHD is a very good approximation everywhere in the magnetosphere. Namely, when plasma inertia and 
temperature are small, the balance of forces acting on the plasma is $\rho {\bf{E}}+{1\over c}{\bf{j}}\times {\bf{B}}=0$, where $\rho$ and $\bf{j}$ are charge and current densities. This simplification allows to solve for the field structure of the magnet
osphere without solving for the plasma dynamics. The first solutions of this kind were found for the axisymmetric aligned rotator. The force-free constraint can be reformulated as an elliptic equation, known as the pulsar equation \cite{michel73a}, which 
describes the equilibrium shape of the magnetic flux surfaces as a function of the poloidal current flowing on them.  This current is not known a priori and can be found through a regularity condition on the light cylinder. The first numerical solution fo
r the magnetopsheric shape was found using an iterative elliptic solver by Contopoulos, Kazanas and Fendt \cite{CKF} (hereafter CKF), and recently reproduced with higher accuracy in \cite{Gruzinov05}.  This solution has a corotating closed zone with rough
ly dipolar field lines extending upto the light cylinder and an open zone, where fieldlines cross the light cylinder, becoming monopolar in the poloidal plane, but increasingly dominated by the toroidal component. This wound-up spiral is carrying Poynting
 flux at the rate $(1\pm 0.1)\mu^2\Omega^4/c^3$ \cite{Gruzinov05}. In order to obtain this very plausible solution, CKF had to assume that the last closed field line extends to the light cylinder. This was used as a boundary condition on the open magnetic
 flux. This assumption was recently relaxed in \cite{Mestel04}, \cite{Contop05} and \cite{Timokhin05}, who find that in principle steady-state solutions with different extent of the closed zone can be obtained. These solutions differ in spindown energy lo
ss and the structure of current sheets and Y-points demarcating the magnetosphere. While steady-state force-free solutions have been invaluable in understanding the pulsar physics, only a time-dependent solution has the potential to remove the ambiguities
 and address the stability and variability questions. 

Below we describe a formulation of time-dependent force-free electrodynamics that we use for simulations (originally derived in \cite{Gruzinov99, Blandford02}). In the presence of strongly magnetized plasma which can short out accelerating fields ($\partial_t ({\bf E}\cdot {\bf B})=0$), the Maxwell equations together with the force-free condition give: 
\begin{eqnarray}
{1\over c} {\partial  {\bf E}\over \partial t}={\bf \nabla} \times {\bf B}&-&{4 \pi \over c} 
{\bf j}; \quad {1\over c} {\partial {\bf B}\over \partial t} =-{\bf \nabla} \times {\bf E}; \label{m1} \\
{\bf j}={c\over 4 \pi} ({\bf \nabla} \cdot {\bf E}) {{\bf E} \times {\bf B} \over B^2}&+&
{c\over 4 \pi} {({\bf B}\cdot {\bf \nabla}\times {\bf B}-{\bf E}\cdot {\bf \nabla}\times {\bf E}){\bf{B}}
\over B^2}.\label{m3}
\end{eqnarray}
Equation \ref{m3} is a prescription for the plasma current that is needed to satisfy the constraints. The two terms have simple meaning: the first term is the part of the current perpendicular to the fieldlines, given by the advection of charge density wi
th the $E\times B$ drift velocity, while the second term is the parallel component of the current which reduces to the parallel component of ${\bf \nabla}\times {\bf B}$ in steady state. The system (\ref{m1}-\ref{m3}) is hyperbolic and  can be integrated 
forward in time. 
In order to numerically solve it we utilize the close similarity to Maxwell equations and use finite-difference time-domain (FDTD) \cite{Yee66} method commonly used in electrical engineering. It allows for dissipationless propagation of EM waves at the ex
pense of introducing numerical dispersion. The nature of FDTD method requires decentering all components of E and B fields. However, to calculate the current we need to know both fields in the same location, and we use linear interpolation for this. The s
econd term in (\ref{m3}) comes from the perfect conductivity constraint, and is relatively cumbersome to calculate numerically because it requires the interpolation of both fields and field derivatives. Instead of using the full equation (\ref{m3}) we sol
ve an even simpler system: we advance Maxwell equations with only the perpendicular component of the current, and then after every timestep we subtract the accumulated $E_{||}$ component from the electric field. This achieves the same purpose of simulatin
g a perfectly conducting plasma as the full expression (\ref{m3}), but is easier to compute. For boundary conditions we set ${\bf E }=- {\Omega} \times {\bf r} \times {\bf B}$ to simulate a rotating conducting sphere. At the present time we do not have re
liable nonreflecting outer boundary conditions, so we have to use large grids to avoid the reflection of transients. 

The initial experiments with time-dependent force-free electrodynamics were reported in \cite{S04}. There we found that a sphere with monopolar magnetic field emits a torsional Alfven wave and settles to an analytic solution with no irregularities at the 
light cylinder. However, a simulation with a dipole magnetic field ran into problems: the solution tried to form a current sheet in the equator and exploded. Spontaneous formation of current sheets is a generic feature of magnetic configurations with init
ially closed field lines and the simulation should be able to handle such sheets. Upon closer examination we found that problems typically appear after electric field in the simulation exceeds the magnetic field. The equations (\ref{m1}-\ref{m3}) do not e
nforce $E<B$ and even though one can start with such a state, the solution may develop regions with $E>B$. In this case the Alfven speed becomes imaginary and leads to exponential instability in the solution. In the equatorial current sheet of the pulsar 
the situation is even more extreme: $E$ starts to exceed $B$ when $B$ approaches zero. This means that the original assumption of magnetic dominance over plasma is violated and one has to restore plasma effects, such as pressure to sustain the current she
et. However, in the framework of force-free electrodynamics these effects are not included, so there is no hope of resolving a stable current sheet on the grid, even due to numerical effects. One way such a current sheet can exist in the simulation is whe
n it is unresolved inside a cell, with a jump in B field over this cell. The FDTD method allows to support such sharp discontinuities without spreading the current sheet. In order to have the method capture such current sheets, we enforce the condition $E
<B$ after every timestep by resetting the electric field in the regions where it grows too large. The physics behind this drastic measure is related to dissipation processes that will happen if $E \times B$ drift approximation is broken and particles acce
lerate while decoupling from magnetic field lines. 

\begin{figure} 
\unitlength = 0.0011\textwidth
\hspace{1\unitlength}
\begin{picture}(280,400)(0,0)

\includegraphics[scale=.32]{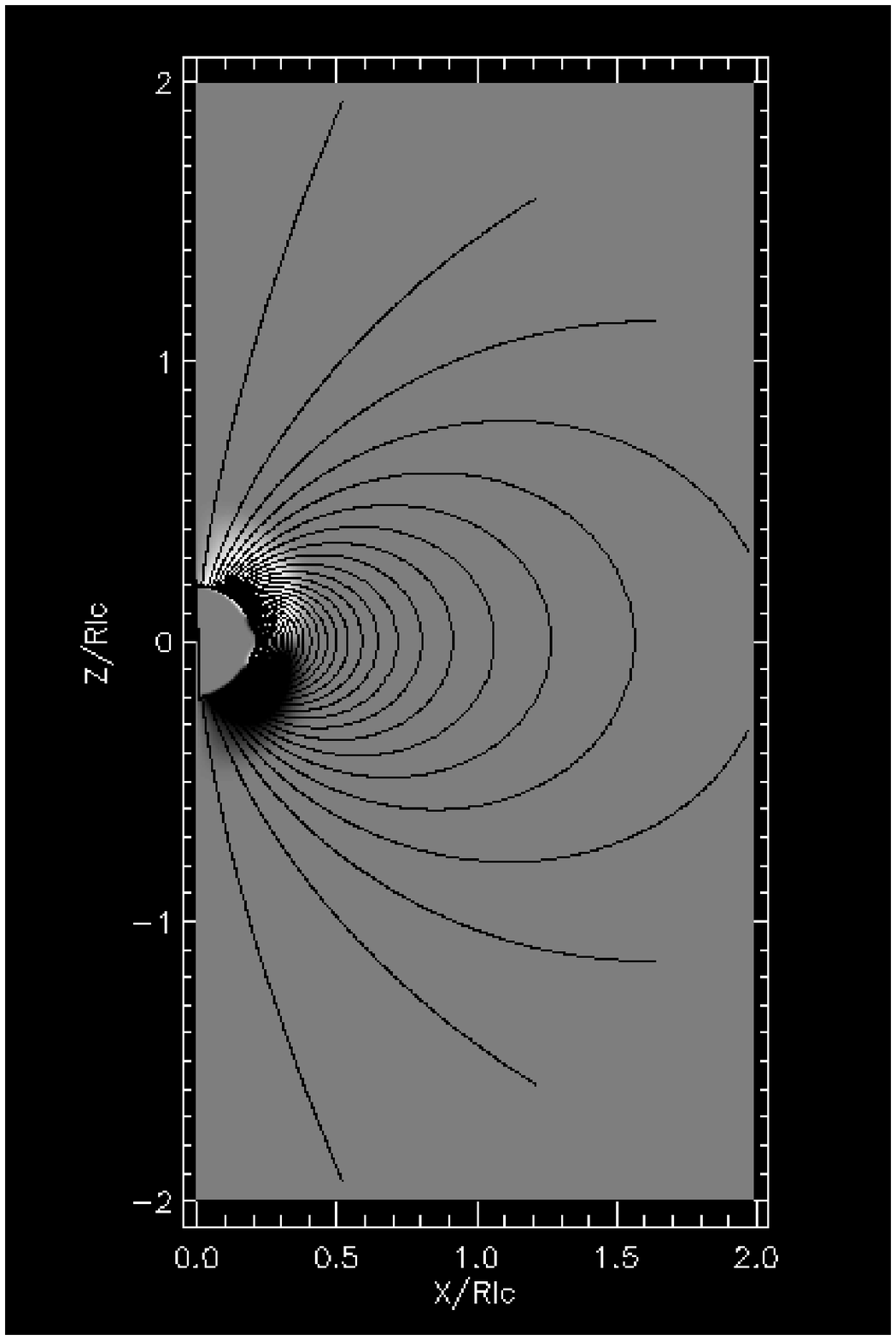}
\end{picture}
\hspace{1\unitlength}
\begin{picture}(280,400)(0,0)
\includegraphics[scale=.32]{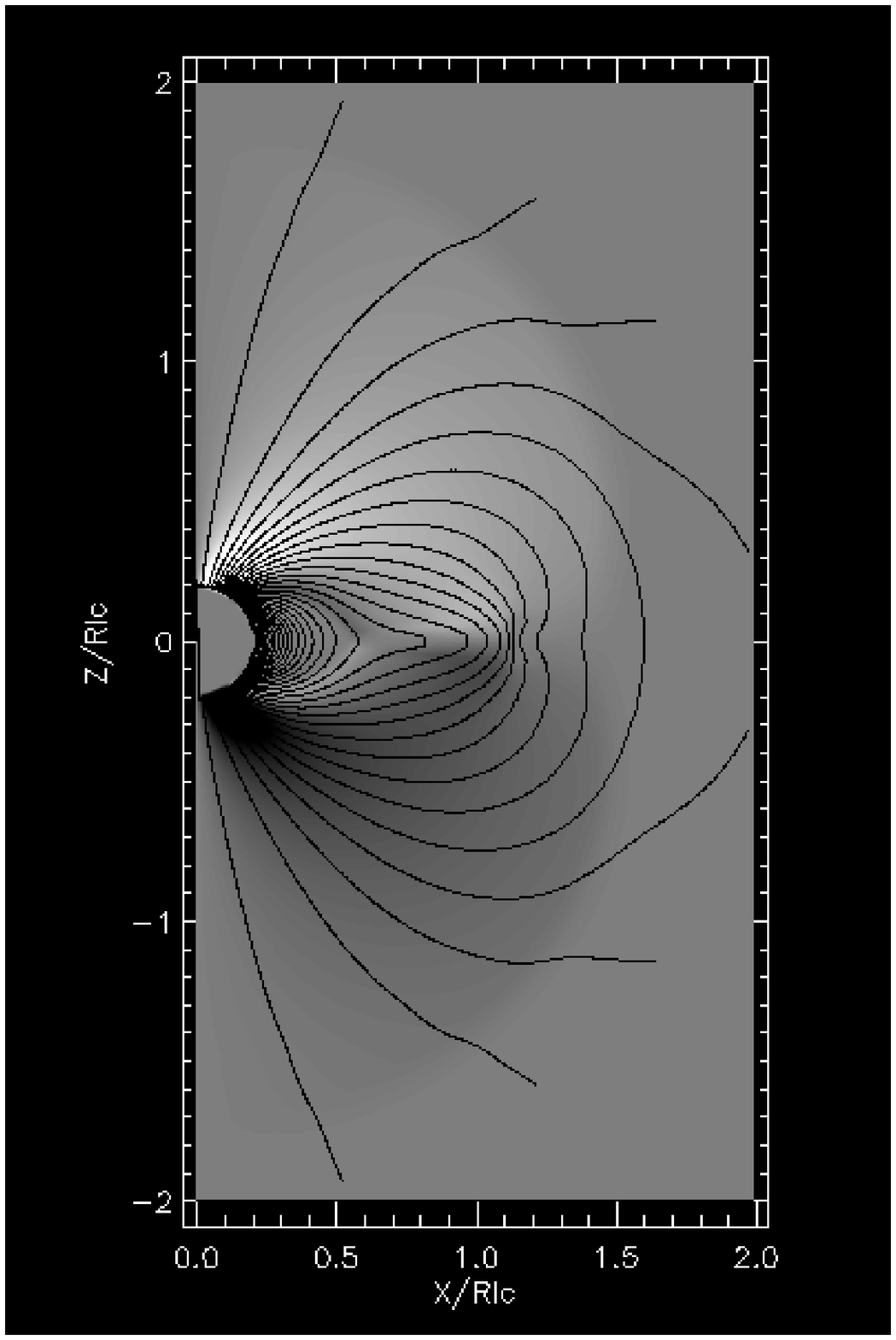} 
\end{picture}
\hspace{1\unitlength}
\begin{picture}(280,400)(0,0)
\includegraphics[scale=.32]{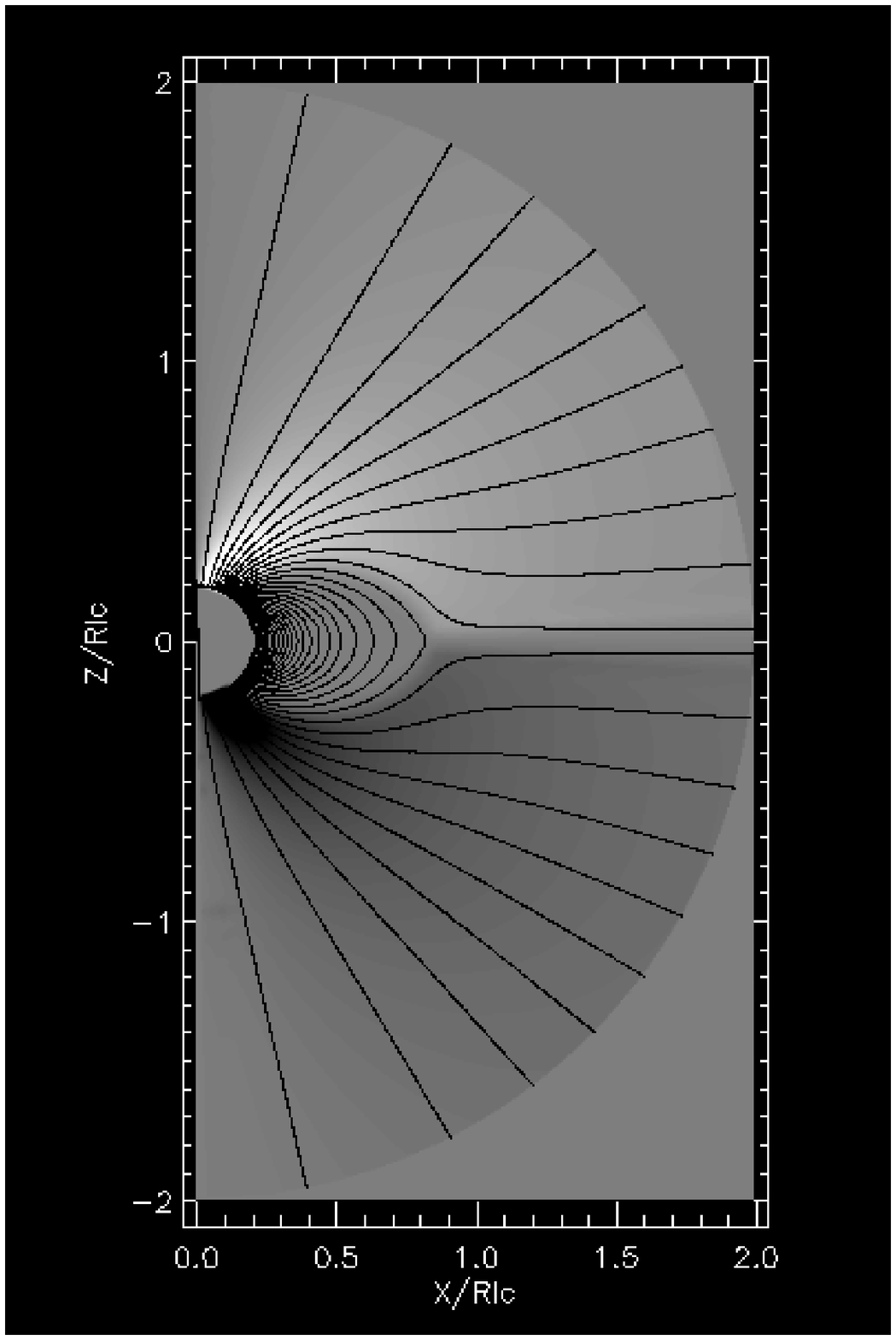}
\end{picture}
\hspace{1\unitlength}
\caption{
Establishment of the rotating aligned dipole magnetosphere. Poloidal field lines are black and the grayscale represents the toroidal magnetic field.}
\label{figevol}
\end{figure}

The sequence of time evolution of the force-free pulsar magnetosphere is shown in figure \ref{figevol}. We begin with a dipole field immersed into conducting inertialess plasma and start rotation of the conducting sphere at the center. A torsional Alfven 
wave propagates from the star, polarizing the space around it and setting the fieldlines (and plasma) into rotation. The current associated with the Alfven wave follows the fieldlines. While initially flowing in the same direction along all fieldlines, th
e current starts to diminish where waves from opposite hemispheres cancel near the equator. This way a set of fieldlines with no net poloidal current starts to grow from the star. On the edge of this zone a return current is flowing which comes from the l
ast uncanceled wave. The rotation of the closed zone carries toroidal current which stretches out the poloidal fieldlines. On the equator the magnitude of the magnetic field hits zero around $0.6 R_{LC}$ after a quarter of stellar rotation. This is where 
the equatorial current sheet first appears. Over the following evolution the fieldlines stretch out further, essentially forming a closed-open configuration of CKF, but with the Y-point well inside the light cylinder. This configuration is not stable, how
ever, and over 5-10 turns of the star some opened field lines snap back and reconnect, moving the Y-point out towards the light cylinder. In the simulations that we have done the Y-point approaches $0.85 R_{LC}$. The magnetic field near the Y-point has an
 interesting discontinuity as pointed out in \cite{Gruzinov05}: the magnitude of magnetic field has a sharp peak near the Y-point before dropping to zero in the current sheet. The field actually has an integrable singularity when $R_Y=R_{LC}$, and this po
int is supported in the code by the action of our artificial resistivity. It is our expectation that the Y-point can get very close to the light cylinder if more realistic resistivity model is used. In our simulations we see saturation of the spindown pow
er of the pulsar at $(1.1 \pm 0.2)\mu^2\Omega^4/c^3$, consistent with time-independent models which place the Y-point at the light cylinder. Thus the time-dependent force-free evolution seems to prefer the maximally extended closed zone, which is consiste
nt with this configuration having the lowest spindown power of other intermediate equilibria. 

We have developed a numerical method for solving time-dependent force-free MHD equations and have applied it to solving a dynamic pulsar magnetosphere. This solution opens new possibilities in application to time-dependent pulsar phenomenology, such as  d
rifting subpulses. It also allows the study of particle acceleration processes in the magnetosphere in the context of a global self-consistent model. The numerical method can also be applied to other differentially rotating magnetized configurations, such
 as magnetospheres of flaring magnetars or coronae of accretion disks. Initial experiments demonstrate the generic tendency of moving fieldlines to spontaneously form current sheets and straighten to break magnetic connection between shearing conductors. 
These reconfigurations are also accompanied by ejection of plasmoids, which could be interesting for dynamics of afterglows. The method can also be extended to 3D to study the spindown of an oblique rotator -- a problem that has escaped solution for 40 ye
ars. 



\bibliographystyle{aipprocl} 



\end{document}